\documentclass[aps,pra,twocolumn,showpacs,superscriptaddress,floatfix]{revtex4}

\usepackage{graphicx}
\usepackage{units}
\usepackage{amsmath,amsfonts}

\renewcommand{\i}{i}
\renewcommand{\d}{\mathrm{d}}
\newcommand{\e}{e}

\newcommand*{\cis}[1]{\e^{\i\,#1}}
\renewcommand{\c}{\mathrm{c}}
\newcommand{\s}{\mathrm{s}}
\newcommand{\R}{\mathrm{R}}
\renewcommand{\L}{{\mathrm{L}}}
\newcommand{\LHM}{\textsc{lhm}}
\newcommand{\RHM}{\textsc{rhm}}
\newcommand{\NLSE}{\textsc{nlse}}
\newcommand{\HSS}{\textsc{hss}}
\newcommand{\MI}{\textsc{mi}}

\begin{document}

\title{Negative diffraction pattern dynamics in nonlinear cavities with left-handed materials}

\author{Pascal Kockaert}
\affiliation{Optique et acoustique, 
             Universit\'{e} libre de Bruxelles CP\,194/5;
             50,~Av.~F.D.~Roosevelt; B-1050 Bruxelles (Belgium)}

\author{Philippe Tassin}
\affiliation{Department of Applied Physics and Photonics, 
             Vrije Universiteit Brussel;
             Pleinlaan 2; B-1050 Brussel (Belgium)}

\author{Guy \surname{Van der Sande}}
\affiliation{Department of Applied Physics and Photonics, %
             Vrije Universiteit Brussel; %
             Pleinlaan 2; B-1050 Brussel (Belgium)}
\affiliation{Optique non lin\'{e}aire th\'{e}orique, %
             Universit\'{e} libre de Bruxelles CP\,231; %
             Campus Plaine, B-1050 Bruxelles (Belgium)}

\author{Irina Veretennicoff}
\affiliation{Department of Applied Physics and Photonics, %
             Vrije Universiteit Brussel; %
             Pleinlaan 2; B-1050 Brussel (Belgium)}

\author{Mustapha Tlidi}
\affiliation{Optique non lin\'{e}aire th\'{e}orique, %
             Universit\'{e} libre de Bruxelles CP\,231; %
             Campus Plaine, B-1050 Bruxelles (Belgium)}

\date{\today}

\begin{abstract}
  We study a ring cavity filled with a slab of a right-handed material
  and a slab of a left-handed material. Both layers are assumed to be
  nonlinear Kerr media. First, we derive a model for the propagation
  of light in a left-handed material. By constructing a mean-field
  model, we show that the sign of diffraction can be made either
  positive or negative in this resonator, depending on the thicknesses
  of the layers. Subsequently, we demonstrate that the dynamical
  behavior of the modulation instability is strongly affected by the
  sign of the diffraction coefficient. Finally, we study the
  dissipative structures in this resonator and reveal the predominance
  of a two-dimensional up-switching process over the formation of
  spatially periodic structures, leading to the truncation of the
  homogeneous hysteresis cycle.
\end{abstract}


\maketitle

\section{Introduction}

In recent years, considerable progress has been realized in the
understanding of dissipative structures in spatially extended systems.
The concept of dissipative structures in systems far from equilibrium
has been first introduced in the context of reaction diffusion
systems~\cite{Glansdorff-1971,Turing-1952}. Their formation is
attributed to the balance between nonlinearities (chemical reaction or
light-matter interaction), transport processes (diffusion and/or
diffraction), and dissipation. Driven optical cavities filled with
nonlinear media -- such as Kerr materials, liquid crystals,
semiconductors, or quadratic media -- belong to this field of research
and constitute a basic configuration in nonlinear optics. It has been
shown, both experimentally and numerically, that these simple devices
support stationary dissipative structures that can be either periodic
or localized in space (for an overview,
see~\cite{Lange-Jun2000,Rosanov-2002,Staliunas-2003,%
  Mandel-Sep2004,Akhmediev-2005}).

Along another line of research, theoretical and experimental studies
have shown the possibility of making left-handed materials (\LHM),
i.e., materials with simultaneously negative permittivity and
permeability. Although the early \LHM{s} were build for microwaves,
the first optical \LHM{s} have been demonstrated
recently~\cite{Moser-Feb2005,Grigorenko-Nov2005,Alu-Feb2006,%
  Linden-May2006,Feth-May2006,Zhang-May2006,Dolling-May2006}.  In his
seminal work, Veselago has shown the possibility of light propagation
in these materials and predicted a number of exotic effects like
negative refraction, inverse Doppler shifts and radiation
tension~\cite{Veselago-1968}. Several experiments have confirmed that
\LHM{}s must be characterized by a negative index of refraction ($n <
0$) and, hence, by a negative wavenumber ($k_0 <
0$)~\cite{Shelby-Apr2001,Wiltshire-Feb2001}.  Linear optical
resonators filled with \LHM{s} were shown to exhibit unique
properties, most noteworthy their unconventional dispersion
relation~\cite{Engheta-Jan2002}. Also, periodic heterostructures that
combine positive and negative refractive indexes present analogies
with cavities and show interesting behavior. In particular, a new
bandgap corresponding to zero averaged refractive index
appears~\cite{Feise-Aug2004,Hegde-Jul2005}.  Despite the importance of
nonlinear phenomena to the propagation of radiation in \LHM{}s, only a
few works have addressed their nonlinear
properties~\cite{Zharov-Jul2003,Agranovich-2004,Feise-2004,%
  Lazarides-Mar2005,Zharov-Aug2005,Shadrivov-Mar2006}.  In particular,
it was shown that \LHM{}s exhibit a bistable behavior with respect to
power, which suggests to use them in power limiters and all-optical
switches.

\begin{figure}[htb]
  \includegraphics[width=8.5cm]{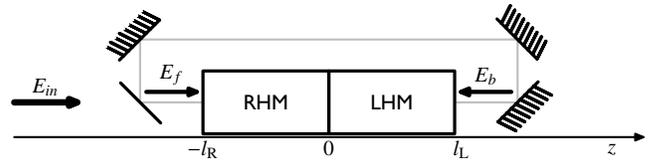}
\caption{\label{Fig:Geometry}
  Geometry of the ring cavity filled with right-handed (\RHM) and
  left-handed (\LHM) materials. The cavity is driven by the external
  field $E_{in}$. The internal forward and backward fields are
  denoted by $E_{f}$ and $E_{b}$, respectively.}
\end{figure}

To the best of our knowledge, the latter studies addressing the
nonlinear behavior of \LHM{s} consider the laser beam as homogeneous
in the plane transverse to the beam propagation axis and therefore
neglect diffraction. The purpose of this article is to report on
analytical and numerical investigations of an externally driven ring
cavity filled with two materials having indices of refraction of
opposite signs, taking diffraction into account (see
Fig.~\ref{Fig:Geometry}).  In a very recent
study~\cite{Gabitov-May2006} a 2-layer (\LHM+\RHM) nonlinear
Fabry-Perot interferometer is considered, but the propagation in the
\LHM{} is assumed to be linear and the study is focused on the
behavior of the device when the input angle is varied. In that case,
surface waves at the interfaces are to be considered, which is not
necessary in our study, where the pump beam is orthogonal to the
entrance surface.

We show that the dynamical behavior of the modulational instability
(\MI) is strongly affected by the left-handed material layer. This is
due to the possibility of controlling the magnitude and even the sign
of the diffraction coefficient by varying the relative thicknesses of
both layers in the resonator. Finally, we describe the two-dimensional
up-switching process that dominates the spatiotemporal behavior of the
nonlinear resonator. Note that the term \textit{negative diffraction}
has been introduced in other systems too, such as atomic Bose-Einstein
condensates \cite{Conti-2004} and discrete
systems~\cite{Christodoulides-1988,Eisenberg-2000}.  Furthermore, the
possibility of diffraction engineering using photonic crystals has
been raised in~\cite{Staliunas-Aug2003}, where it was shown that the
width of solitons may become as small as a few times the wavelength of
the modulation period. We want to propose a completely different mechanism
for tuning the diffraction coefficient. In addition, the solitons of
Ref.~\cite{Staliunas-Aug2003} were studied with only one transverse
dimension, while here, we consider \mbox{1-D} and \mbox{2-D} systems
and we show that beyond the similarity of their evolution equations,
the one- and two-dimensional cases exhibit very different dynamics.

In this article, we consider an optical ring cavity, which is driven
by a coherent light beam of frequency $\nu_0=\omega_0/2\pi$. The
cavity is filled with two adjacent layers containing a right-handed
material (\RHM) and a left-handed material (\LHM), respectively. We
will assume that both materials are local, dispersive and have a Kerr
type nonlinearity in their dielectric behavior. These assumptions are
quite common for traditional (\RHM) materials, but need some
explanation for \LHM{s}. First, the absence of nonlocality can be
justified from the fact that \LHM{s} are subwavelength-structured
materials; coupling between resonators over distances longer than the
wavelength will therefore be negligible with respect to diffraction.
Second, \LHM{}s have been shown to be dispersive in both their
dielectric and magnetic response~\cite{Veselago-1968}. The third
assumption seems to contradict the magnetic nonlinearity found by
Zharov~\textit{et.~al.}~\cite{Zharov-Jul2003}. However, their
nonlinearity is a consequence of a nonlinear dielectric inserted
between the gaps of the split-ring resonators in Smith's metamaterial,
forming nonlinear capacitors. This will be unlikely the case for
optical \LHM{s}. If the nonlinearity can be introduced by other means,
e.g. by using nonlinear resonators interacting with the electric
field~\cite{Lapine-2003}, the nonlinearity can indeed be modelled by a
third-order susceptibility for a centrosymmetric medium. Furthermore,
it can be inferred from the analysis presented below that a Kerr
nonlinearity in the \RHM{} suffices to observe the effects reported in
this paper and that the \LHM{} may even be simply linear.

The structure of this paper is as follows. In
Sect.~\ref{Sect:Propagation}, we will derive a nonlinear Schr\"{o}dinger
equation describing the propagation of light through the \LHM.
Subsequently, in Sect.~\ref{Sect:Mean Field}, a mean-field model for
the nonlinear resonator will be constructed. A linear stability
analysis of this model will be carried out in
Sect.~\ref{Sect:Stability} and the dissipative structures emerging due
to the modulational instability will be discussed in
Sect.~\ref{Sect:Discussion}.

\section{Light Propagation in a \LHM}\label{Sect:Propagation}
Light coupled into the resonator depicted in Fig.~\ref{Fig:Geometry}
will propagate through the right-handed layer as well as through the
left-handed layer. A physical model of this resonator should therefore
include a description of both interactions. 

On the one hand, it is well-known that propagation through a
right-handed Kerr type nonlinear material can be properly treated with
the standard nonlinear Schr\"{o}dinger equation
(\NLSE)~\cite{Moloney-2003}, which gives the envelope $A$ of the
electric field as a function of time $t$, the longitudinal coordinate
$z$, and the transverse coordinates $x$ and $y$:
\begin{equation}
  \i\frac{\partial A}{\partial \zeta} 
+ \frac{c}{2 \omega_0 n_0} \nabla^2_\perp A 
- \frac{\beta_2}{2} \frac{\partial^2 A}{\partial t^2} 
+ \frac{3 \omega_0}{2 c n_0} \chi_0^{(3)} \left|A\right|^2 A
=0, \label{Eq:NLSE-RHM}
\end{equation}
where $\zeta$ is the longitudinal coordinate of a reference system traveling with
the group velocity along the cavity, $\omega_0$ is the central
pulsation of the pulse envelope,  $n_0$ is the refractive index
of the material at \(\omega_0\), $\chi_0^{(3)}$ its third-order nonlinear
susceptibility at \(\omega_0\), and \(c\) is the speed of light in
vacuum (\(\varepsilon_0\mu_0\,c^2=1\)). 

On the other hand, propagation through a nonlinear left-handed
material has not been considered before. It is not obvious at all that
the \NLSE{} is still valid for left-handed materials, because it is
commonly derived for nonmagnetic materials with
$\mu_r=\mu/\mu_0\equiv1$.  Therefore, in this Section we will outline
how the \NLSE{} can be generalized for dispersive magnetic media. A
detailed treatment can be found in Ref.~\cite{Tassin-Aug2005}.

In the first step, we model the material behavior by the following
constitutive equations relating the displacement field $\mathbf{D}$
and the electric field $\mathbf{E}$ on the one hand, and between the
magnetic field $\mathbf{H}$ and the induction field $\mathbf{B}$ on
the other hand:
\begin{gather}
  \mathbf{D}(\mathbf{r},t) 
= \varepsilon_0 \int_{-\infty}^{+\infty}  \varepsilon_r(t-\tau)
                                        \mathbf{E}(\mathbf{r},\tau) \d\tau 
+  \mathbf{P}^{(3)}, 
\label{Eq:ConstEqD} \\
\begin{split}
  \mathbf{P}^{(3)} 
= \varepsilon_0 \int \int &\int_{-\infty}^{\infty} 
                           \chi^{(3)}(t-\tau_1,t-\tau_2,t-\tau_3) \\
                          &E(\mathbf{r},\tau_1) 
                           E(\mathbf{r},\tau_2)
                           \mathbf{E}(\mathbf{r},\tau_3) 
                \d\tau_1 \d\tau_2 \d\tau_3,
\end{split}\label{Eq:ConstEqNL} \\
\mathbf{B}(\mathbf{r},t) 
= \mu_0 \int_{-\infty}^{+\infty}
        \mu_r(t-\tau)\mathbf{H}(\mathbf{r},\tau) \d\tau 
\label{Eq:ConstEqB}.
\end{gather}
These equations account for the dispersive dielectric and magnetic
behavior by the convolution integrals and for the Kerr nonlinearity by
the third-order polarization field $\mathbf{P}^{(3)}$. The equations
are valid for dispersive magnetic materials that are cubic and
centrosymmetric. Combining
Eqs.~(\ref{Eq:ConstEqD})-(\ref{Eq:ConstEqB}) with Maxwell's equations,
we obtain the wave equation governing the evolution of the electric field
in a nonlinear dispersive magnetic dielectric:
\begin{widetext}
\begin{equation}
-\nabla \times \nabla \times \mathbf{E}(\mathbf{r},t) 
-\frac{1}{c^2} \int_{-\infty}^{\infty} 
                  \mu_r(t-\tau)
                  \,\frac{\partial^2}{\partial \tau^2}
                  \int_{-\infty}^{\infty}
                  \varepsilon_r(\tau-u)\,\mathbf{E}(\mathbf{r},u) 
                \,\d{u}\d\tau 
= \mu_0 \int_{-\infty}^{\infty} 
                  \mu_r(t-\tau) 
                  \,\frac{\partial^2}{\partial \tau^2}
                  \mathbf{P}^{(3)}(\mathbf{r},\tau) 
        \,\d\tau 
\label{Eq:WaveEq}
.
\end{equation}
\end{widetext}
One can now proceed with the derivation of an amplitude equation for
quasi-monochromatic light by invoking the slowly-varying envelope
approximation and by applying the classical multiple scales
perturbation technique~\cite{Moloney-2003}. In this derivation, we
represent the coherent light beam by the following wavepacket
\begin{equation}
\mathbf{E} =\mathbf{e} \; A(\mathbf{r},t) \e^{\i(\mathbf{k}_0\cdot
  \mathbf{r}-\omega_0 t)} + \mathrm{c.c.}, 
\label{Eq:WavePacket}
\end{equation}
and we obtain a generalized version of Eq.~(\ref{Eq:NLSE-RHM}) that is
valid for both right-handed and left-handed materials:
\begin{equation}
\i\frac{\partial A}{\partial \zeta} 
+ \frac{c}{2 \omega_0 n_0} \nabla^2_\perp A 
- \frac{\beta_2}{2} \frac{\partial^2 A}{\partial t^2} 
+ \frac{3 \omega_0}{2 c \eta_0} \chi_0^{(3)} \left|A\right|^2 A=0
\label{Eq:NLSE-LHM}
,
\end{equation}
with \(\eta_0=\sqrt{\varepsilon_r/\mu_r}\), the characteristic
impedance of the medium.  We thus find that the evolution of the
envelope of the electric field is still described by a \NLSE, but the
coefficients of Eq.~(\ref{Eq:NLSE-LHM}) differ from those of
Eq.~(\ref{Eq:NLSE-RHM}). In particular, the refractive index $n_0$ in
the nonlinear term of the original equation must be replaced by the
characteristic impedance $\eta_0$.  From our detailed analysis, it is
now also clear that the coefficient of the diffraction term is still
inversely proportional to the refractive index, and therefore negative
in a \LHM{}.

\section{Resonator Model}\label{Sect:Mean Field}

Diffraction in the ring cavity depicted in Fig.~\ref{Fig:Geometry}
acts with the usual sign in the \RHM, and can be compensated for by
negative diffraction in the \LHM. In order to get a proper
understanding of the dynamics arising from the presence of the \LHM\
layer in the cavity, we implement the procedure of Lugiato and Lefever
to build a mean-field model (or LL-model) describing the evolution of
the field in the cavity~\cite{Lugiato-May1987}. This model is based on
three additional assumptions about the optical cavity. First, we
consider the \RHM\ and the \LHM\ to be adequately impedance matched,
so that reflections at their interfaces can be neglected. As a result,
the backward field ($E_{b}$) is negligible. Note that this
condition could also be imposed in the system, by adding an isolator
in the ring cavity. Second, we assume that the aspect ratio of the
cavity is very large (high Fresnel number). In this case, the
boundary conditions in both transverse directions do not affect the
central part of the beam. Third, we consider cavities that are shorter
than the diffraction and the nonlinearity space scales.

For ease of notation, the time coordinate $t$ has been normalized with
respect to the lifetime of a photon in the cavity. This results in the
slow dimensionless time scale $\tau = \sqrt{2}\pi t / \left(\mathcal{F}
  T\right)$, where $T$ is the cavity round-trip
time and $\mathcal{F}$ its finesse. The transverse space coordinates
$x$ and $y$ and the longitudinal space coordinate $z$ are left
unchanged. With these conventions, the field circulating in the ring
cavity can be described by $E_{f} = A(x,y,\tau) \exp{\left[
    {\i\left( {k_0 z - \omega_0 t} \right)} \right]}$, with $k_0$ the
wavenumber. The evolution of the envelope $A(x,y,\tau)$ is then governed
by (see Appendix)
\begin{equation}
\frac{\partial A}{\partial \tau}
 =
-(1+\i\Delta)A + \mathcal{E} + \i \Gamma {\left| A \right|}^2 A + \i {\delta} \nabla_\bot^2 A,
\label{eq:LL}
\end{equation}
where $\nabla_\bot^2$ is the transverse Laplace operator. The
parameter $\Delta$ in Eq.~(\ref{eq:LL}) is the detuning and is related
to the linear phase accumulated by the light during one round-trip,
$\psi = - 2\pi\Delta/\mathcal{F}$. The driving field is taken to be $E_{in}
= 2\pi \mathcal{E} / \mathcal{F}  \exp({-\i \omega_0 t})$ and the
coefficient of the nonlinear term describes the combined effect of the
nonlinearities in both material layers: $\Gamma = \mathcal{F}\left(
  {\gamma_\R\ l_\R + \gamma_\L l_\L} \right)/(2\pi)$, where the
subscripts $\R$ and~$\L$ denote material parameters in the \RHM\ and
\LHM, respectively. A self-focusing nonlinearity ($\Gamma > 0$) is
assumed in the remainder of this paper; the case of a self-defocusing
medium can be studied in a similar way. Finally, the diffraction
coefficient ${\delta}$ is given by
\begin{equation}
{\delta} 
= 
\frac{\mathcal{F}}{2\pi}\left( {\frac{l_\R}{2 k_\R} + \frac{l_\L}{2 k_\L}} \right) = \frac{\lambda_0 \mathcal{F}}{8\pi^2} \left( {\frac{l_\R}{n_\R} - \frac{l_\L}{\left| {n_\L} \right|}} \right).
\label{eq:diffrcoeff}
\end{equation}

Equation~(\ref{eq:LL}) is similar to the well-known LL-model. However,
there is an essential difference: the diffraction coefficient can
become negative. From Eq.~(\ref{eq:diffrcoeff}), we deduce that this
occurs when the length~$l_\L$ of the \LHM\ layer is larger than $-k_\L
l_\R / k_\R$. We will demonstrate below that this negative diffraction
may have dramatic consequences on the pattern formation process.

\section{Linear Stability Analysis}\label{Sect:Stability}

The homogeneous steady state solutions (\HSS) $A_\s$ of
Eq.~(\ref{eq:LL}) are $\mathcal{E} = [1+\i(\Delta-\Gamma\,\vert A_\s
\vert^2)] A_\s$. The intracavity intensity $\vert A_\s \vert^2$ as a
function of 
$\vert\mathcal{E}\vert^2$ is single-valued for $\Delta < \sqrt{3}$ and
multiple-valued for $\Delta > \sqrt{3}$. We performed the stability
analysis of these \HSS. With transverse periodic boundaries, the
deviation from the steady state is taken proportional to
$\exp{\left(\i \,\mathbf{k} \cdot \mathbf{r} + \lambda \tau \right)}$
with $\mathbf{k} = (k_{\mathrm{x}},k_{\mathrm{y}})$ and
$\mathbf{r}=(x,y)$. Setting $\lambda = 0$ leads to a $k$-dependent
marginal stability condition:
\begin{equation}
1 + \left( {\Delta + {\delta} k^2 - 2 \Gamma \left| {A_\s} \right|^2} \right)^2 = \Gamma^2 \vert A_\s \vert^4.
\end{equation}
The condition ${\partial \lambda }/{\partial k} = 0$ then yields the
expression of the threshold associated with the modulational
instability (often called Turing instability): 
$\vert A_\c \vert 
= 
1/\sqrt{\Gamma}$ and 
$\Gamma\vert\mathcal{E}_\c\vert^2 
= 
\Delta^2-2\Delta+2$. At
this bifurcation point, the critical wavelength is
\begin{equation}
\Lambda_\c 
= 2\pi\sqrt{\frac{{\delta}}{2-\Delta}} 
= \sqrt{\frac{\pi \mathcal{F} 
  \left( {k_\L l_\R + k_\R l_\L} \right)}{k_\R k_\L \left( {2 - \Delta} \right)}}.
\label{eq:wavelength}
\end{equation}

\begin{figure}[thb]
  \includegraphics[width=8cm]{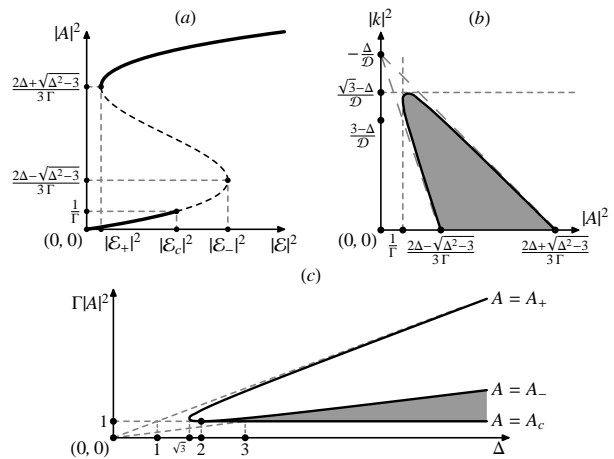}
\caption{\label{Fig:BistaInsta}
  (a) Bistability curve of the intracavity intensity
  ($\vert{}{A}\vert^{2}$) vs driving intensity ($\vert \mathcal{E}
  \vert^2$).
  (b) Marginal stability curve. The domain of instability
  is filled. 
  (c) Intracavity intensity vs detuning ($\Delta$). The plane is
  divided by the curves $A = A_\pm$ and $A = A_\c$, with 
  $\vert{A_\c}\vert^2 =  1/\Gamma$ and 
  $\vert{A_\pm}\vert^2 = \left( 2\Delta \pm
    \sqrt{\Delta^2-3}\right) / (3\Gamma)$. The domain of \MI\ 
  is filled.}
\end{figure}

If the thickness of the \LHM\ layer is small, $l_\L < -k_\L l_\R /
k_\R$, the system will have a positive diffraction coefficient
(${\delta} > 0$) and will behave like any traditional nonlinear
ring cavity. The more interesting situation occurs when $l_\L > -k_\L
l_\R / k_\R$. In that case, the diffraction coefficient is negative,
i.e., ${\delta}<0$. The \HSS\ are unaltered, but their stability is
completely different from the positive diffraction case. The upper
branch is now stable whatever the value of $\Delta$. This means that
the presence of a \LHM\ completely inhibits the \MI\ in the
monostable regime ($0 < \Delta < \sqrt{3}$). In contrast, for
$\Delta > 2$, an additional portion of the lower branch exhibits \MI,
i.e., when $1/\Gamma < \vert A \vert^2 < \vert A_- \vert^2$, or
equivalently, when $\mathcal{E}_\c < \mathcal{E} < \mathcal{E}_-$.
The intensities 
$\vert A_\pm \vert^2 
= 
(2\Delta \pm \sqrt{\Delta^2-3} ) / (3\Gamma)$
corresponding to the driving fields 
$\mathcal{E}_\pm 
= 
\sqrt{|A_\pm|^2 [1+(\Delta-|A_\pm|^2)^2]}$ are the
coordinates on the bistability curve of the two limit points of the
homogeneous hysteresis cycle. The behavior of the negative diffraction
resonator is illustrated in Fig.~\ref{Fig:BistaInsta} for 
$\Delta > 2$. In Fig.~\ref{Fig:BistaInsta}(a), we have plotted the \HSS\ as a
function of the driving field; the domain of instability is indicated
by dashed lines. Figure~\ref{Fig:BistaInsta}(b) shows the marginal
stability curve. In Fig.~\ref{Fig:BistaInsta}(c), the threshold
associated with the \MI\ as a function of the detuning parameter is
plotted. From the latter figure, it is clear that the stability domain
is bounded and that only a small portion of the lower \HSS\ branch is
affected by the \MI. This is in strong contrast with traditional
resonators where the upper \HSS\ branch is unstable and where the
instability domain is unbounded.

\begin{figure}[tbh]
  \includegraphics[width=8cm]{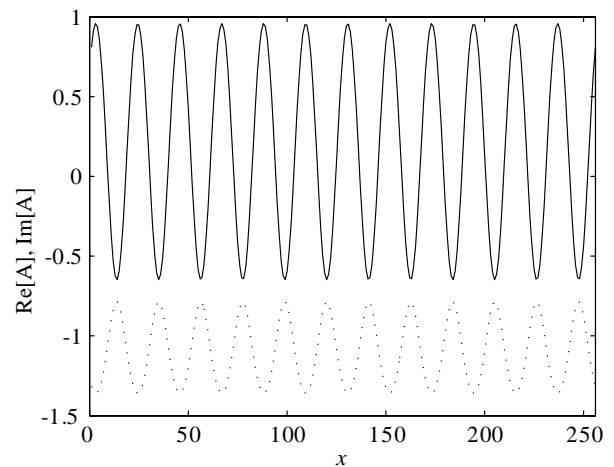}
\caption{\label{Fig:PatternOneD}
  Stable dissipative structure emerging in the one-dimensional case
  when the resonator is biased near the threshold of \MI. The real and
  imaginary parts of $A$ are plotted with solid and dotted lines,
  respectively. Parameters are $\Delta = 10$, ${\delta} = -2$,
  $\Gamma = 1$ and $\mathcal{E} = 9.25$.}
\end{figure}

\section{Discussion}\label{Sect:Discussion}

When the driving field amplitude is in the range $\mathcal{E}_\c <
\mathcal{E} <\mathcal{E}_-$ and $\Delta > 2$, a finite band of
linearly unstable Fourier modes $k$ exist [see Fig.~2(b)], which
trigger the spontaneous evolution of the intracavity field amplitude
towards a stationary, spatially periodic distribution that occupies
the whole cavity space available in the transverse directions. An
example of a dissipative structure in the one-dimensional system
obtained by numerical simulation of Eq.~(\ref{eq:LL}) is shown in
Fig.~\ref{Fig:PatternOneD}. The initial condition consists of some
small random noise added to the \HSS\ and periodic boundary conditions
are assumed. To compare with the analytical results, we calculate the
wavelength ($\Lambda_\c = 3.14$) given by the linear stability
analysis [Eq.~(\ref{eq:wavelength})] and the wavelength of the stable
dissipative structure obtained numerically ($\Lambda_\mathrm{num} =
3.23$). A very good agreement between the two wavelengths is obtained.
A weakly nonlinear analysis performed in the vicinity of the threshold
for \MI\ shows that the branch of periodic solutions always appears
subcritically for $\Delta > \frac{41}{33}$. Changing the diffraction
coefficient in these simulations will simply change the scaling of the
pattern, as can be seen from Eq.~(\ref{eq:LL}) by nondimensionalizing
$x$ and $y$ using the transformation $(x,y) =
\sqrt{{\delta}}(X,Y)$. Since we can make the diffraction
coefficient of the resonator under study as small as we desire by
choosing the thicknesses of the two material layers appropriately, we
can in principle produce dissipative structures with sizes smaller
than the diffraction-limited size $\sqrt{l \mathcal{F} / (2 \pi
  k_0)}$. The combined use of \LHM{s} and \RHM{s} therefore enables us
to circumvent the diffraction limit, which must be compared to the
sub-diffraction-limited imaging of \LHM\ plates~\cite{Pendry-2000}.

\begin{figure}[htb]
\includegraphics[width=\linewidth]{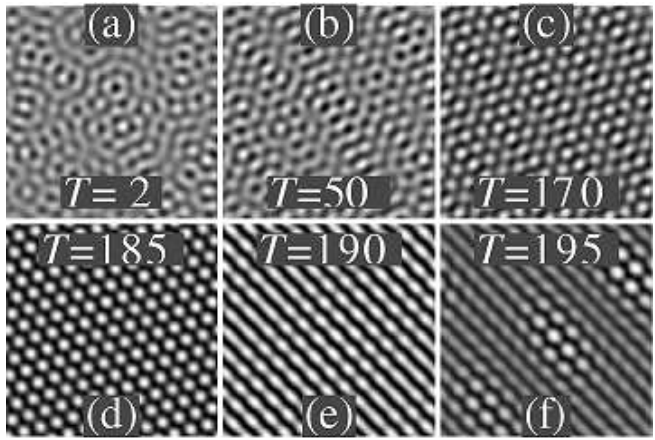}
\caption{\label{Fig:Patterns}
  Dynamics of the pattern formation in the two-dimensional case.
  (a)--(b) Initial pattern formation. (c)--(d) Hexagonal structures
  emerge. (e) Hexagonal structures are replaced by stripes. (f)
  Self-organized spots appear. The labels correspond to the points
  (a)--(f) on Fig.~\ref{Fig:PatEvol}. The parameters are the same as
  those in Fig.~\ref{Fig:PatternOneD}. Maxima are plain white and the
  integration mesh is $256 \times 256$.}
\end{figure}
\begin{figure}[htb]
\includegraphics[width=7cm]{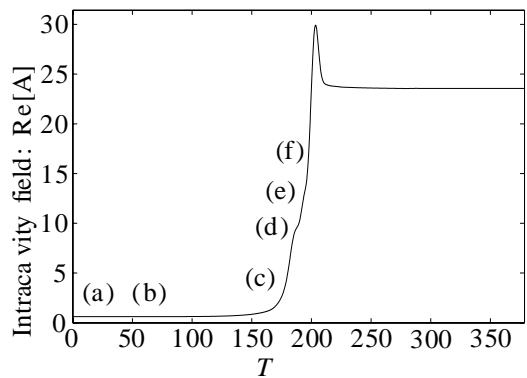}
\caption{\label{Fig:PatEvol}
  Time evolution during the two-dimensional up-switching process
  described. The average maximum value of the real part of the
  intracavity field is plotted.}
\end{figure}

We carried out similar simulations in a square-shaped domain using
periodic boundary conditions.  The two-dimensional patterns are
obtained by perturbing initially the \HSS\ in the same parameter range
as in Fig.~\ref{Fig:PatternOneD}. The following sequence of transverse
patterns is obtained: hexagons, stripes, and self-organized spots.
This behavior is shown in Fig.~\ref{Fig:Patterns}(a)--(f), where we
have plotted the real part of the intracavity field during the
spatiotemporal evolution. From Fig.~\ref{Fig:PatEvol}, we see that the
average intensity grows exponentially, which is clearly due to the
amplification of unstable wavelengths around $\Lambda_\c = 2\pi/k_c$,
as can be expected from the linear stability analysis. Finally, the
field amplitude reaches the high \HSS. Note that the transition
between the two \HSS\ is abrupt, but takes place after a fairly long
transient close the lower \HSS\ branch. By varying the input field
amplitude, we found that this phenomenon occurs for all input fields
in the range $\mathcal{E}_\c < \mathcal{E} < \mathcal{E}_-$. This
spontaneous spatiotemporal up-switching process therefore leads to the
truncation of the homogeneous hysteresis cycle.

To summarize, we presented a description of light propagation in a
ring cavity filled with a left-handed and a right-handed material,
both materials exhibiting a Kerr type nonlinearity. We derived a
mean-field model for the spatiotemporal evolution of this resonator
and showed that its diffraction coefficient can become negative if the
thickness of the \LHM\ layer is large enough. In this case, the
\LHM\ stabilizes the upper branch of the homogeneous steady state
curve, and destabilizes a small part of the lower branch for high
detuning ($\Delta > 2$). In the latter parameter range, we
demonstrated that pattern formation can take place. However, a
spontaneous two-dimensional up-switching process dominates the
spatiotemporal dynamics over the formation of stable patterns,
resulting in the truncation of the homogeneous hysteresis cycle.

\begin{acknowledgments}

This work was supported by the Belgian Science Policy Office under
grant IAP-V18 (Belgian Photon Network). P.\ T.\ is a Research
Assistant of the Research Foundation - Flanders (\textsc{FWO} -
Vlaanderen). M.\ T.\ is a Research Associate of the Fonds national de
la recherche scientifique (\textsc{FNRS}, Belgium).

\end{acknowledgments}

\appendix*

\section{Mean-field model}

The mean-field approximation has already been applied to systems with
boundary conditions~\cite{Tlidi-Mar2000}. Here, the propagation model
is different, and we detail how it is possible to simplify the system
of four coupled propagation equations, and the boundary conditions to
only one equation of evolution, similar to the equation of the
LL-model~\cite{Lugiato-May1987}.

\subsection{Hypothesis and notations}

We consider bidirectional propagation in a ring cavity with impedance
matching at the interfaces between the two layers of classical and
left-handed materials. In order to neglect the backward field with
respect to the forward field in certain expressions, we also assume
that the reflections at the other interfaces are limited. In cases
where this assumption would not be verified, it is possible to add a
circulator in the cavity, so that the condition holds.

Below, the letters \(R\), \(L\), and \(C\) will refer to the field,
nonlinearity and dispersion in, respectively , the right-handed
medium, the left-handed medium, and the cavity outside of the two
layers. The quantities \(A\) and \(B\) are (resp.)  the envelope of
the forward and the backward fields travelling in the ring cavity.

With theses notations, for example, \(A_{C}(-l_R)\) is the
electric field outside the right-handed medium, in the entrance plane,
while \(A_\R(-l_R)\) is the same field, at the same interface, but inside
the medium.

The mean-field approximation consists in approximating the solution of
the evolution in each layer, and for each forward or backward
component as follows. In order to simplify the derivation, we rewrite
Eq.~(\ref{Eq:NLSE-LHM}) in the form
\begin{equation}
\frac{\partial A}{\partial \zeta}
=
\i\left[{\mathcal{D}}+\mathcal{N}(A,B)\right]\,A,
\label{Eq:NLS-operators}
\end{equation}
where \({\mathcal{D}}\) is a linear differential operator, and
\(\mathcal{N}(A,B)\) a nonlinear operator, i.e.
\begin{eqnarray}
{\mathcal{D}}&=&
 \frac{1}{2 k_0}\nabla^2_\perp-\frac{\beta_2}{2}\frac{\partial^2}{\partial{t}^2},\\
\mathcal{N}&=&\frac{3\omega_0}{2c\eta_0}\chi_0^{(3)}
              \left(
                       \left\vert{A}\right\vert^2
                +\sigma\left\vert{B}\right\vert^2
              \right) A
 ,
\end{eqnarray}
where \(\sigma\) is the nonlinear coupling coefficient between the
forward and the backward waves.

\subsection{Boundary conditions}

The geometry of the ring cavity leads to the following boundary conditions,
\begin{eqnarray}
A_{C}(-l_R)   &=&t_{in}\,E_{in}  + r_{in}\,\cis{\varphi}\,A_{C}(l_L)\label{eq:BC:a},\\
A_R(-l_R) &=&t_{R}\,A_{C}(-l_R) + r_{R}\,B_{R}(-l_R)\label{eq:BC:b},\\
A_L(0)    &=&A_R(0)\label{eq:BC:c},\\
A_{C}(l_L)    &=&t_{L}\,A_L(l_L) + r_{in}\,r_{L}\,\cis{\varphi}\,B_{C}(-l_R)\label{eq:BC:d}.
\end{eqnarray}
These expressions define the transmission and reflection coefficients
at the interfaces. The phase factor \(\cis{\varphi}\) accounts for the
total phase accumulated in the ring cavity, outside of the two layers.

Similar relations can be written for the backward field,
\begin{eqnarray}
B_{C}(-l_R)   &=&t_{R}\,B_{R}(-l_R)  + r_{R}\,A_{C}(-l_R)\label{eq:BC:e},\\
B_L(0)    &=&B_R(0)\label{eq:BC:f},\\
B_L(l_L)  &=&t_{L}\,B_{C}(l_L) + r_{L}\,A_{L}(l_L)\label{eq:BC:g},\\
B_{C}(l_L)    &=&r_{in}\,\cis{\varphi}\,B_{C}(-l_R)\label{eq:BC:h}.
\end{eqnarray}

\subsection{Approximate equations of evolution}

Assuming that \(\delta\zeta=O(1)\), a first order integration of
Eq.~(\ref{Eq:NLS-operators}) leads to
\begin{eqnarray}
A(\zeta+\delta\zeta)&=&A(\zeta)+\i\delta\zeta
           \left[
             {\mathcal{D}}+\mathcal{N}(A,B)
           \right]\vert_\zeta\,A+O(2)\nonumber\\
          &=&\e^{\i\delta\zeta
           \left[
             {\mathcal{D}}+\mathcal{N}(A,B)
           \right]\vert_\zeta}\,A+O(2)\nonumber\\
          &=&\e^{\i\delta\zeta{\mathcal{D}}\vert_\zeta}
             \e^{\i\delta\zeta\mathcal{N}(A,B)\vert_\zeta}
             \,A+O(2)\nonumber\\
          &=&\e^{\i\delta\zeta\mathcal{N}(A,B)\vert_\zeta}
             \e^{\i\delta\zeta{\mathcal{D}}\vert_\zeta}  
             \,A+O(2)\nonumber.\\\label{eq:evol}
\end{eqnarray}
Now, we make the assumption that the reflection coefficients are small
quantities (\(r_{L}=O(1)=r_{R}\)), and that the backward field is initially
very weak (\(B=O(1)\)) in comparison with the forward field, so that
\begin{eqnarray}
A_{C}(-l_R)   &=&t_{in}\,E_{in}  + r_{in}\,\cis{\varphi}\,A_{C}(l_L)\label{eq:BC:O2:a},\\
A_R(-l_R) &=&t_{R}\,A_{C}(-l_R) + O(2)\label{eq:BC:O2:b},\\
A_L(0)    &=&A_R(0)\label{eq:BC:O2:c},\\
A_{C}(l_L)    &=&t_{L}\,A_L(l_L) + O(2)\label{eq:BC:O2:d},\\
B_{C}(-l_R)   &=&O(0)\,B_{R}(-l_R)  + O(1)\,A_{C}(-l_R)\label{eq:BC:O2:e},\\
B_L(0)    &=&B_R(0)\label{eq:BC:O2:f},\\
B_L(l_L)  &=&O(0)\,B_{C}(l_L) + O(1)\,A_{L}(l_L)\label{eq:BC:O2:g},\\
B_{C}(l_L)    &=&O(1)\,B_{C}(-l_R)\label{eq:BC:O2:h}.
\end{eqnarray}
In addition, the first order integration of the propagation equation Eq.~(\ref{eq:evol}) shows that \begin{eqnarray}
A_{R}(0)    &=& O(0) A_{R}(-l_R),\\
A_{L}(l_L)  &=& O(0) A_{L}(0),\\
B_{R}(-l_R) &=& O(0) B_{R}(0),\\
B_{L}(0)    &=& O(0) B_{L}(l_L).\\
\end{eqnarray}
From Eqs.~(\ref{eq:BC:O2:a})--(\ref{eq:BC:O2:h}), we deduce that 
\begin{eqnarray}
A &=& O(0) E_{in},\\
B &=& O(1) A.\\
\end{eqnarray}
This shows that the backward field will stay a first order quantity if
it is initially so with respect to the forward
field, and that we can neglect the linear coupling with the backward
field in the boundary conditions
Eqs.~(\ref{eq:BC:O2:a})--(\ref{eq:BC:O2:d}), or
Eqs.~(\ref{eq:BC:a})--(\ref{eq:BC:d}), as the coupling is of second order.

The nonlinear coupling between the forward and the backward field is
incoherent, meaning that it depends only on the intensity of the
fields, and involves no exchange of energy between the two
counterpropagating envelopes. This allows to relate, at first order, the
forward and backward intensities everywhere in the cavity to the
forward input intensity.

A reduction of Eqs.~(\ref{eq:BC:a})--(\ref{eq:BC:d}) and (\ref{eq:evol}) provides
\begin{eqnarray}
\lefteqn{A^{(n)}(-l_R)=}\\
 && t_{in}\,E_{in} + r_{in}\,t_L\,t_R\,\cis{\varphi}\,\times\nonumber\\
 && \cis{l_L\mathcal{N}_L}\cis{l_L{\mathcal{D}}_L}\,
             \cis{l_R{\mathcal{D}}_R}\cis{l_R\mathcal{N}_R}\,
             A^{(n-1)}(-l_R),
\nonumber
\end{eqnarray}
which expresses how the envelope of the forward field is related to
both this same envelope one roundtrip before and the driving field
\(E_{in}\).

This discrete infinite set of equations can be reduced to a single
equation, by introducing the roundtrip time \(T\), and approximating
the discrete variations by a partial derivative with respect to the
slow time \(\tau=n\,T\):
\begin{equation}
  A^{(n)}-A^{(n-1)}\simeq T \left.\frac{\partial{A}}{\partial{\tau}}\right\vert_{\tau=n\,T}.
\end{equation}
We find directly that 
\begin{equation}
  T \frac{\partial{A}}{\partial{\tau}}
 =
  t_{in}\,E_{in} + \left(\rho\cis{\mathcal{P}}-1\right) A,
\label{eq:A:continuous}
\end{equation}
where
\(\mathcal{P}=l_L{\mathcal{D}}_L+l_R{\mathcal{D}}_R+l_L\mathcal{N}_L+l_R\mathcal{N}_R\)
accounts for the total propagation, and \(\rho=r_{in}\,t_L\,t_R\cis{\varphi}\).

The backward field is proportional to the forward field, with
\(\vert{B_{C}(-l_R)}\vert= \vert{r_R+r_L t_L t^2_R}\vert
\vert{A_{C}(-l_R)}\vert +O(2)\), and similar expressions for
\(\vert{B_R(-l_R)}\vert=\theta_R\vert{A_R(-l_R)}\vert\), and
\(\vert{B_L(0)}\vert=\theta_L\vert{A_R(0)}\vert\). The exact
expression of \(\theta_R\) and \(\theta_L\) is not important to
complete our derivation. Below, it is only needed to remember that
\(\theta=O(1)\), and that \(\theta\) is real.

The linear relation between the forward and the backward intensities
simplifies the calculation of the nonlinear operators, as
\begin{equation}
\mathcal{N}(A,B)
=\gamma\left(\vert{A}\vert^2+\sigma\vert{B}\vert^2\right)
=\gamma\left(1+\sigma\theta^2\right)\,\vert{A}\vert^2,
\end{equation}
with \(\gamma=\frac{3\omega_0}{2c\eta_0}\chi_0^{(3)}\).

By redefining \(\gamma^\prime=\left(1+\sigma\theta^2\right)\gamma\), we
can completely hide the existence of the backward field, that appears
as a second order correction to the nonlinear coefficient.  Therefore
\begin{eqnarray}
\mathcal{P}&=&\frac{1}{2}\left(\frac{l_L}{k_L}    + \frac{l_R}{k_R}     \right) 
                       \nabla^2_\perp 
            - \frac{1}{2}\left(l_L \beta_{2,L}     + l_R \beta_{2,R}     \right) 
                       \frac{\partial^2}{\partial{t}^2}\nonumber\\
           &+&           \left(l_L \gamma_L^\prime + l_R \gamma_R^\prime \right) 
                       \left\vert{A}\right\vert^2\label{eq:P}
.
\end{eqnarray}

\subsection{Lugiato-Lefever equation}

Together, equations (\ref{eq:A:continuous}) and~(\ref{eq:P}) fully describe the
dynamics of the cavity. In this section, we will assume that we are
working near resonance of the cavity, so as to obtain the standard
form of the LL-model, after proper normalization of the coefficients.

Near resonance, \(\rho=\vert\rho\vert\cis{\psi}\) is real, and
\(\psi\) accounts for the detuning with respect to the resonance
(\(\psi\approx0\)).  In this particular case,
\begin{equation}
\rho\cis{\mathcal{P}}
=\vert\rho\vert\cis{(\psi+\mathcal{P})}
=\vert\rho\vert\left(1+\i\psi+\i\mathcal{P}\right)+O(2).
\end{equation}
Introducing this expression in Eq.~(\ref{eq:A:continuous}), we find
\begin{equation}
T \frac{\partial{A}}{\partial\tau}
=
t_{in} E_{in}
+
\left[
  \left(\vert\rho\vert-1\right)+\i\vert\rho\vert\psi+\i\vert\rho\vert\mathcal{P}
\right]\,A.
\end{equation}
This equation is similar to the LL-model, which appears more clearly, if we redefine
\begin{eqnarray}
\tau^\prime      &=&  \frac{1-\vert\rho\vert}{T}\,\tau,\label{eq:redim:a}\\
\mathcal{E}_{in} &=&  \frac{t_{in}}{1-\vert\rho\vert}\,E_{in},\label{eq:redim:b}\\
\Delta          &=&  -\frac{\vert\rho\vert}{1-\vert\rho\vert}\,\psi,\label{eq:redim:c}\\
\delta          &=&   \frac{1}{2}
                      \frac{\vert\rho\vert}{1-\vert\rho\vert}
                      \left(\frac{l_L}{k_L} + \frac{l_R}{k_R} \right),\label{eq:redim:d}\\
\beta           &=&   \frac{\vert\rho\vert}{1-\vert\rho\vert}
                      \left(l_L \beta_{2,L} + l_R \beta_{2,R} \right),\label{eq:redim:e}\\
\Gamma          &=&   \frac{\vert\rho\vert}{1-\vert\rho\vert}
                       \left(l_L (1+\sigma\theta_L^2) \gamma_L 
                           + l_R (1+\sigma\theta_R^2) \gamma_R \right),\nonumber\\
\label{eq:redim:f}
\end{eqnarray}
as in this case, the dynamics in the cavity near resonance is described by
\begin{equation}
\frac{\partial{A}}{\partial\tau^\prime}
=
\mathcal{E}_{in}
-(1+\i\Delta)A
+\i\delta\nabla^2_\perp A
-\i\frac{\beta}{2} \frac{\partial^2A}{\partial{t}^2}
+\i\Gamma\vert{A}\vert^2 A.
\label{eq:LL-extended}
\end{equation}

When the system is driven by a continuous wave, the dispersive term
disappears and we obtain Eq.~(\ref{eq:LL}). It is convenient to introduce
the finesse of the cavity
\(\mathcal{F}=2\pi\frac{\vert\rho\vert}{1-\vert\rho\vert}\). If we neglect the
second-order quantities \(\theta\) and assume that
\(\vert\rho\vert\approx{}\vert{r_{in}}\vert=t_{in}=1/\sqrt{2}\),
Eqs.~(\ref{eq:redim:a})--(\ref{eq:redim:f}) lead to the simple form
\begin{eqnarray}
\tau^\prime      &=&  \frac{\sqrt{2}\pi}{\mathcal{F}T}\,\tau,\label{eq:redim:F:a}\\
\mathcal{E}_{in} &=&  \frac{\mathcal{F}}{2\pi}\,E_{in},\label{eq:redim:F:b}\\
\Delta          &=&  -\frac{\mathcal{F}}{2\pi}\,\psi,\label{eq:redim:F:c}\\
\delta          &=&   \frac{\mathcal{F}}{4\pi}
                      \left(\frac{l_L}{k_L} + \frac{l_R}{k_R} \right),\label{eq:redim:F:d}\\
\Gamma          &=&   \frac{\mathcal{F}}{2\pi}
                       \left(l_L  \gamma_L 
                           + l_R  \gamma_R \right)\label{eq:redim:F:f}
.
\end{eqnarray}


\newcommand{\singleletter}[1]{#1}

\end{document}